 \documentclass[twocolumn,showpacs,groupedaddress]{revtex4}

\usepackage{graphicx}
 
\newcommand{\epsfigbox}[5]{%
\begin{figure} \vspace{#3}%
\includegraphics{#2}%
\caption{ 
\label{fig:#1} #5} 
\vspace{#4} 
\end{figure}} 
 
\newcommand{\epsfigwide}[5]{%
\begin{figure*} \vspace{#3}%
\includegraphics{#2}%
\caption{ 
\label{fig:#1} #5} 
\vspace{#4} 
\end{figure*}} 
 
\begin{document}   
 
\title{Nature of excited 0$^+$ states in $^{158}$Gd 
described by the projected shell model} 
\author{Yang Sun$^{1,2}$,  
        Ani Aprahamian$^{1}$, Jing-ye Zhang$^{3}$, Ching-Tsai Lee$^{3}$} 
\affiliation{   
$^1$Department of Physics, University of Notre Dame, Notre Dame, 
Indiana 46556, USA\\  
$^2$ Department of Physics, Xuzhou Normal University, Xuzhou, 
Jiangsu 221009, P. R. China\\
$^3$Department of Physics and Astronomy, University of Tennessee, Knoxville, 
Tennessee 37996, USA  
}   
   
\date{\today}   
   
\begin{abstract}   
Excited 0$^+$ states are studied  
in the framework of the projected shell model, aiming at understanding 
the nature of these states in deformed nuclei in general, and  
the recently observed 13 excited 0$^+$ states in $^{158}$Gd in particular. 
The model, which contains projected   
two- and four-quasiparticle states as building blocks  
in the basis,  
is able to reproduce reasonably well the energies for all the  
observed 0$^+$ states. The obtained B(E2) values   
however tend to suggest that these 0$^+$ states might have a mixed 
nature of quasiparticle excitations coupled to collective vibrations.  
\end{abstract}   
   
\maketitle   
   
 
Dynamic perturbations of nuclear shapes around the equilibrium can give rise  
to physical states at low to moderate excitation energies.  
Classical examples of such motion are $\beta$ and $\gamma$ vibrations  
\cite{BM75,Ia84}, in which nucleons undergo vibrations in a collective manner.  
Traditionally, the first excited $K^\pi = 0^+$ states  
and the first excited $2^+$ states are interpreted, respectively,  
as the $\beta$ and $\gamma$ vibrational states. While the $2^+$  
collective excitations are better understood theoretically,  
the nature of the lowest $0^+$ excitation of deformed nuclei  
still remains under debate \cite{CB94,Burke,Kumar,Bonn,Garrett}. 
The physics of  
higher $0^+$ states is even more complex because, on one side,  
they can predominantly be multi-phonon states based on the single-phonons  
\cite{Ani02},  
and on the other side, they can be quasiparticle (qp) excitations in nature.  
The real situation is, perhaps, that the two aspects, collective excitations  
and qp states, are mixed by residual interactions. 
  
Data on $K^\pi = 0^+$ states have been relatively sparse.  
In a very recent work by Lesher {\it et. al.} \cite{Exp}, a remarkable $(p,t)$  
experiment revealed a total of 13 excited 0$^+$ states in $^{158}$Gd,  
below an excitation energy of approximately 3.1 MeV. This abundance  
of 0$^+$ states in a single nucleus  
provides significant new information on the poorly  
understood phenomenon, which has immediately sparked off theoretical interest.  
For this energy range, one may think about an explanation  
through collective modes. In fact, Zamfir, Zhang, and Casten \cite{GCM}  
suggested that many of the observed 0$^+$ states may be of two-phonon  
octupole character. Nevertheless, these authors warned also that,  
although the mechanism was excluded in their collective models,  
many of the 0$^+$ states in this excitation energy range may be  
predominantly two-qp in character. 
   
The purpose of the present paper is to investigate whether one can  
explain the nature of the observed $0^+$ states in terms of qp excitations.  
In contrast to Ref. \cite{GCM}, our calculation here does not emphasize  
the aspect of collective excitation. Although, similar to the work of Zamfir,  
Zhang, and Casten \cite{GCM}, we may provide only a partial image,  
our results may shed light on the importance of considering  
the qp aspects in understanding the nature of these $0^+$ states. 
 
Our study is based on the projected shell model (PSM) \cite{review}. 
The PSM is the spherical shell model built  
on a deformed basis. The PSM calculation usually   
begins with the deformed Nilsson single-particle states at a deformation  
$\varepsilon$.   
Pairing correlations are incorporated into the Nilsson states by BCS   
calculations. The consequences of the Nilsson-BCS calculations provide  
us with a set of qp states that define the qp vacuum  
$\left|\phi(\varepsilon)\right>$.  
One then constructs the shell model bases by building  
multi-qp states. The broken symmetry in these states is  
recovered by angular momentum projection \cite{review}  
(and particle number projection, if necessary) to   
form a shell model basis in the laboratory frame.   
Finally a two-body shell model Hamiltonian  
is diagonalized in the projected space.   
 
To determine the deformation at which the shell model basis should be built,  
we first study the bulk properties of $^{158}$Gd including the deformation.  
In a microscopic calculation, one searches for energy minima by  
varying the deformation parameter. Here, we calculate the  
angular-momentum-projected energies having the form 
\begin{equation}   
E^I(\varepsilon) = {{\left< \phi(\varepsilon)\right|\hat H \hat P^I \left| 
\phi(\varepsilon)\right>} \over {\left< \phi(\varepsilon)\right| \hat P^I 
\left| \phi(\varepsilon)\right>}}, 
\label{energy}   
\end{equation}   
where $P^I$ is the angular-momentum projection operator \cite{RS80} which  
projects the mean-field vacuum $\left|\phi(\varepsilon)\right>$ onto states 
with good angular momentum. As many previous PSM calculations for  
the rare earth nuclei, particles in     
three major shells ($N=4,5,6$ for neutrons and $N=3,4,5$ for protons) are   
activated in the present calculation for $^{158}$Gd.  
For comparison, unprojected energies  
\begin{equation}   
E(\varepsilon) = {{\left< \phi(\varepsilon)\right|\hat H \left|
\phi(\varepsilon)\right>} \over {\left< \phi(\varepsilon) | \phi 
(\varepsilon)\right>}} . 
\label{meanfield}   
\end{equation}   
are also calculated.  
 
As one can see in Fig. 1, the lowest energy for a given angular momentum  
$I$ is well localized at deformations varying from $\varepsilon \approx 0.24$  
at $I=0$ to $\varepsilon \approx 0.28$ at $I=12$.  
The ground-state deformation calculated by M\"oller {\it et. al.} \cite{MN95} 
for this nucleus yields $\varepsilon = 0.25$. 
These angular-momentum-projected minima lie at deformations that are  
slightly larger than the mean-field minimum ($\varepsilon \approx 0.23$).  
Fig. 1 indicates  
that $^{158}$Gd is a stably deformed nucleus against  
rotation, with a pronounced energy minimum corresponding to a  
deformed prolate shape. In the following calculations, we thus  
construct the shell model basis at the deformation $\varepsilon=0.26$.  
The vacuum state $\left| \phi(\varepsilon=0.26) \right>$ is hereafter  
written as $\left|0 \right>$.  
    
\epsfigbox{fg 1}{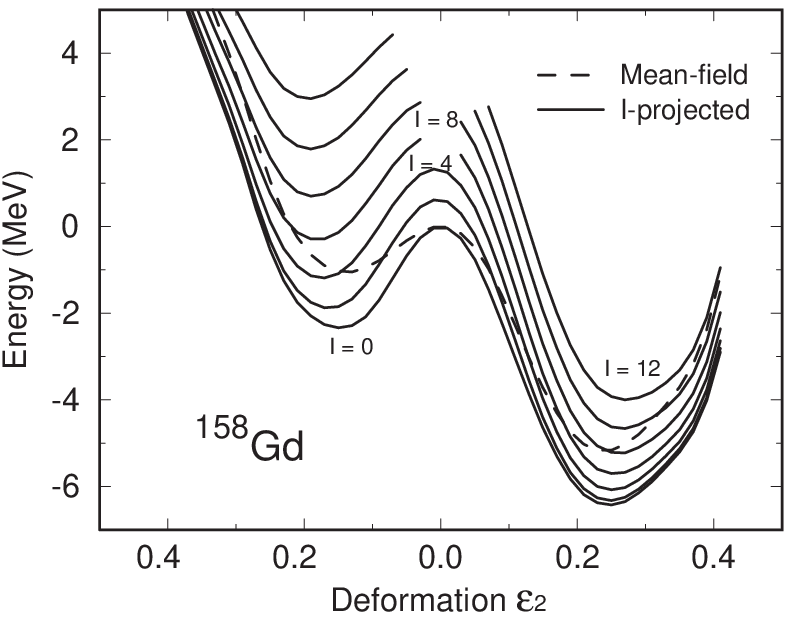}{0pt}{0 pt}{ 
Angular-momentum-projected energies in $^{158}$Gd for the states with  
$I=0$ to $I=12$ as a function of deformation. The unprojected non-rotating 
energies (denoted as mean-field) are also shown for comparison.} 
 
Following the spirit of the Tamm-Dancoff method \cite{RS80}, we build the 
shell model space by including 0-, 2- and 4-qp states:   
\begin{equation}   
\left|\Phi_\kappa\right> = 
\left\{\left|0 \right>, \    
\alpha^\dagger_{n_i} \alpha^\dagger_{n_j} \left|0 \right>,\   
\alpha^\dagger_{p_i} \alpha^\dagger_{p_j} \left|0 \right>,\   
\alpha^\dagger_{n_i} \alpha^\dagger_{n_j} \alpha^\dagger_{p_i}   
\alpha^\dagger_{p_j} \left|0 \right> \right\} ,   
\label{baset}   
\end{equation}   
where $\alpha^\dagger$ is the creation operator for a qp and the   
index $n$ ($p$) denotes neutron (proton) Nilsson quantum numbers  
which run over   
the low-lying orbitals. Thus, the projected multi-qp 
states are the building blocks of our shell model basis: 
\begin{equation}   
\left|\Psi^I_M\right> = \sum_\kappa f^I_\kappa \hat P^I_{MK} 
\left|\Phi_\kappa\right> . 
\label{wavef} 
\end{equation} 
Here, $\kappa$ labels the basis states and  
$ f^I_\kappa$ are determined by configuration mixing.  
   
We then diagonalize the Hamiltonian in the projected multi-qp  
states given in (\ref{wavef}). In the calculation, we employ a quadrupole  
plus pairing Hamiltonian,  
with inclusion of quadrupole-pairing term \cite{review}  
\begin{equation}   
\hat H = \hat H_0 - {1 \over 2} \chi \sum_\mu \hat Q^\dagger_\mu   
\hat Q^{}_\mu - G_M \hat P^\dagger \hat P - G_Q \sum_\mu \hat   
P^\dagger_\mu\hat P^{}_\mu,   
\label{hamham}   
\end{equation}   
where $\hat H_0$ is the spherical single-particle Hamiltonian which   
contains a proper spin-orbit force.    
The quadrupole-quadrupole  
interaction strength $\chi$ is determined by the self-consistent   
relation with deformation $\varepsilon$.  
The monopole pairing strength $G_M$ is taken to be   
$G_M=\left[20-13(N \mp Z)/A\right]/A$ with ``$-$" sign for neutrons and  
``$+$" sign for   
protons.    
Finally, the quadrupole pairing   
strength $G_Q$ is assumed to be proportional to $G_M$, the   
proportionality constant   
being fixed to 0.20 in the present work.   
These interaction strengths are the same    
as the values used in the previous PSM calculations   
for the rare earth nuclei \cite{review}.   
 
\epsfigwide{fg 2}{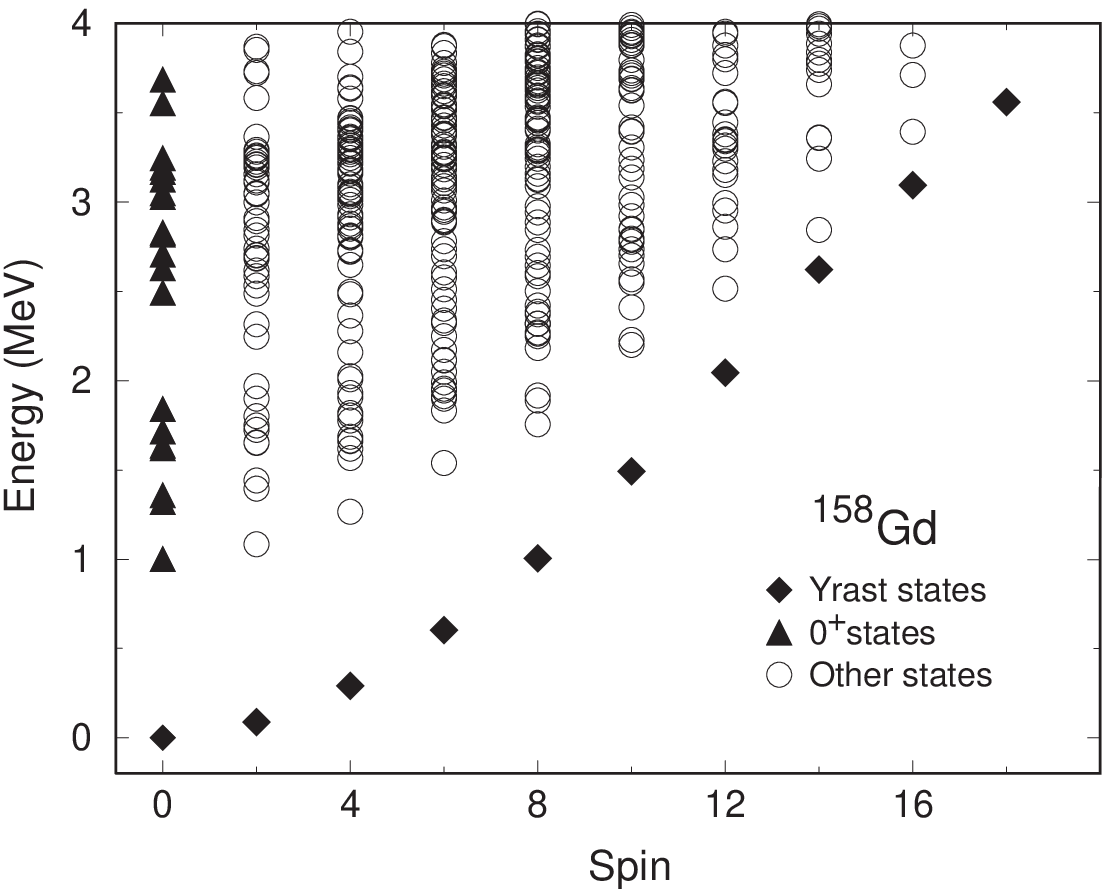}{0pt}{0 pt}{
Theoretical energy spectrum of $^{158}$Gd calculated up to $E=4$ MeV 
and $I=18 \hbar$. Diamonds are the yrast states and filled triangles 
are excited 0$^+$ states.} 
 
In Fig. 2, we show the partial theoretical spectrum in $^{158}$Gd  
up to 4 MeV in energy and  
$18 \hbar$ in spin. We emphasize that all these states have been obtained by a  
single diagonalization, without any adjustment for individual states.  
Out of these many states, let us concentrate on the lowest one at each spin  
(the yrast band, denoted by diamonds), and on all the $0^+$ states  
(denoted by filled tri-angles).  
Although it is not a focus of our discussion in this paper,  
we believe that  
a comparison of the yrast band with known data can provide a strict  
test of the model and can provide 
a useful constraint to the calculations of the $0^+$ states.  
   
\epsfigbox{fg 3}{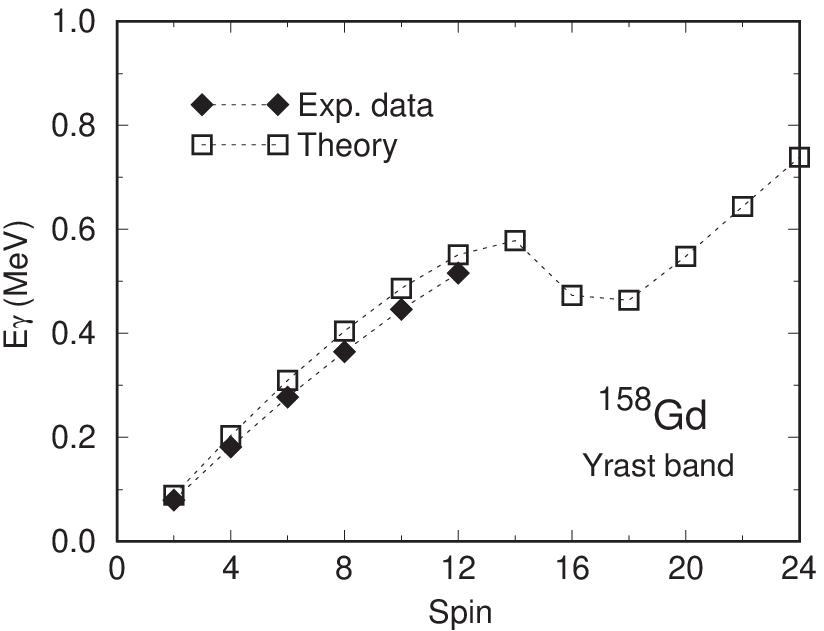}{0pt}{0 pt}{ 
Comparison of the PSM calculations and data for the yrast states in  
$^{158}$Gd in the form of $\gamma$-ray energy   
$E_\gamma (I) = E(I)-E(I-2)$   
versus spin $I$.   
(Data are taken from Ref.    
\protect\cite{isotopes}).} 
 
Fig. 3 presents the PSM results for the yrast band in $^{158}$Gd  
(the same values  
shown in Fig. 2 as diamonds), which are compared with 
the known data \cite{isotopes}, in a plot of $\gamma$-ray energy versus  
spin. As can be seen, the data are described very well. 
The calculations predict a sudden drop in the curve at spin $I=16$,  
corresponding to a backbending in the moment of inertia.  
This sudden change occurs just at the upper part of the band 
where the current measurement stops. Extension from the current measurement 
should see this phenomenon and will provide a strict test of our model  
prediction. 
In this regard, we note that in the isotonic chain of nuclei, $^{160}$Dy, 
$^{162}$Er, $^{164}$Yb, and $^{166}$Hf, similar backbending effects have been 
observed and successfully described elsewhere by the PSM 
\cite{backb1,backb2}.  
   
\epsfigbox{fg 4}{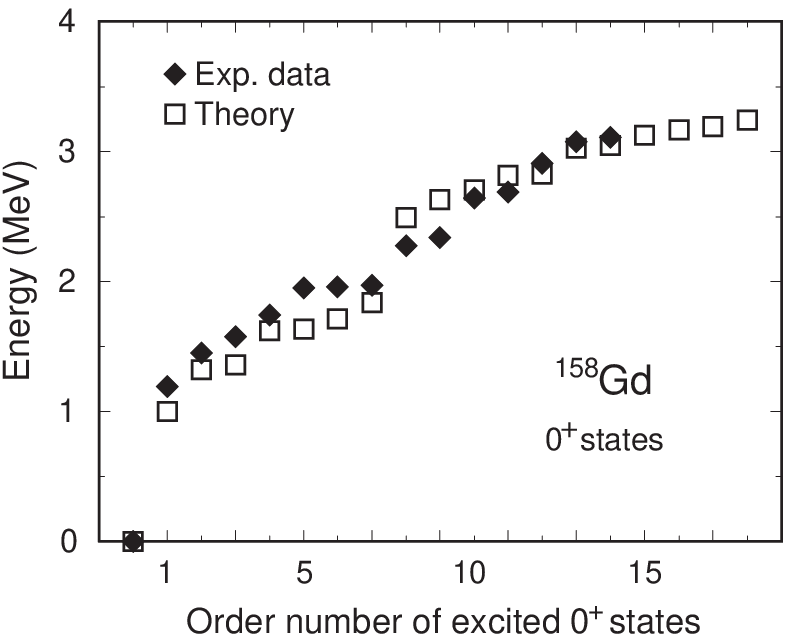}{0pt}{0 pt}{ 
Comparison of the calculated $0^+$ states in $^{158}$Gd with data 
\protect{\cite{Exp}}.} 
 
We now turn our discussion to the excited $0^+$ states.  
Lesher {\it et. al.} \cite{Exp} observed 13 $0^+$ states in the nucleus 
$^{158}$Gd, 
within an excitation energy range of 1.2 MeV to 3.1 MeV.  
In Fig. 4, we plot all the theoretical $0^+$ states (up to 3.2 MeV in 
energy; the same values 
shown in Fig. 2 as filled triangles) in the order of excitation energies. 
The experimental data \cite{Exp} are shown in the same plot for comparison. 
The predicted $0^+$ states are found to be in the right energy range, 
although deviations  
can be clearly seen between theory and experiment.  
 
What we found impressive is the number of $0^+$ states predicted by the 
calculation. The PSM produces a sufficient number of $0^+$ states to be 
compared with data. As described in Eq. (\ref{wavef}), the total 
wave function $\left|\Psi^I_M\right>$ is a linear combination of the 
(projected) basis states given in (\ref{baset}).   
The basis states in (\ref{baset}) are not arbitrarily selected but are taken 
from all the neutron and proton Nilsson orbitals that lie close to the 
Fermi surface. In $^{158}$Gd, the relevant orbitals are:  
${3\over 2}[521]\nu$, ${5\over 2}[523]\nu$, ${11\over 2}[505]\nu$, 
${3\over 2}[651]\nu$, and ${5\over 2}[642]\nu$ for neutrons, 
and ${1\over 2}[411]\pi$, ${3\over 2}[411]\pi$, ${5\over 2}[413]\pi$, 
${5\over 2}[532]\pi$, and ${7\over 2}[523]\pi$ for protons.  
In each of these 10 near-Fermi orbitals, nucleons having opposite signs 
for the $K$ quantum number can couple to a 2-qp state with total $K=0$. 
Combination of a pair of 2-qp states can further give $K=0$ 4-qp states. 
If one neglects the coupling of these qp states to the collective states,  
the number of low-lying $0^+$ states   
in deformed nuclei obtained in this way depend solely  
on the single-particle level density and the  
level distribution near the Fermi surface.   
Since similar conditions can also be found in many other 
rare earth 
nuclei, we expect such an abundance of $0^+$ states as found in $^{158}$Gd  
not to be an isolated case. We predict that such an abundance of 
$0^+$ states should also be observed in many other nuclei. 
 
\begin{table*}  
\caption{Predicted 2-qp and 4-qp $0^+$ states (below 3.25 MeV) in $^{158}$Gd. 
}  
\begin{tabular}{|c|c|c|c|}  
\hline  
E (MeV)&B(E2, $0^+_i \rightarrow 2^+_g$) (W.u.)&qp-states & Configurations \\  
\hline  
1.004 & 1.87  & 2-qp & $-{5\over 2}[642]\nu, {5\over 2}[642]\nu$ \\ 
1.321 & 0.004 & 2-qp & $-{11\over 2}[505]\nu, {11\over 2}[505]\nu$ \\ 
1.360 & 0.014 & 2-qp & $-{3\over 2}[411]\pi, {3\over 2}[411]\pi$ \\ 
1.621 & 0.076 & 2-qp & $-{5\over 2}[413]\pi, {5\over 2}[413]\pi$ \\ 
1.636 & 0.041 & 2-qp & $-{3\over 2}[521]\nu, {3\over 2}[521]\nu$ \\ 
1.716 & 0.234 & 2-qp & $-{5\over 2}[523]\nu, {5\over 2}[523]\nu$ \\ 
1.841 & 1.102 & 2-qp & $-{5\over 2}[532]\pi, {5\over 2}[532]\pi$ \\ 
2.492 & 0.328 & 2-qp & $-{7\over 2}[523]\pi, {7\over 2}[523]\pi$ \\ 
2.631 & 0.001 & 4-qp & $-{11\over 2}[505]\nu, {11\over 2}[505]\nu,  
                -{3\over 2}[411]\pi, {3\over 2}[411]\pi$ \\ 
2.707 & 0.161 & 2-qp & $-{3\over 2}[651]\nu, {3\over 2}[651]\nu$ \\ 
2.818 & 0     & 4-qp & $-{11\over 2}[505]\nu, {11\over 2}[505]\nu,  
                -{5\over 2}[413]\pi, {5\over 2}[413]\pi$ \\ 
2.829 & 0     & 4-qp & $-{3\over 2}[521]\nu, {3\over 2}[521]\nu, 
                -{3\over 2}[411]\pi, {3\over 2}[411]\pi$ \\ 
3.028 & 0     &4-qp & $-{3\over 2}[521]\nu, {3\over 2}[521]\nu,  
                -{5\over 2}[413]\pi, {5\over 2}[413]\pi$ \\ 
3.048 & 0     & 4-qp & $-{5\over 2}[523]\nu, {5\over 2}[523]\nu, 
                -{3\over 2}[411]\pi, {3\over 2}[411]\pi$ \\ 
3.126 & 0.002 & 2-qp & $-{1\over 2}[411]\pi, {1\over 2}[411]\pi$ \\ 
3.168 & 0     & 4-qp & $-{11\over 2}[505]\nu, {3\over 2}[521]\nu,  
                 {5\over 2}[413]\pi, {3\over 2}[411]\pi$ \\ 
3.192 & 0     & 4-qp & $-{3\over 2}[521]\nu, -{5\over 2}[523]\nu, 
                 {5\over 2}[413]\pi, {3\over 2}[411]\pi$ \\ 
3.245 & 0     & 4-qp & $-{5\over 2}[523]\nu, {5\over 2}[523]\nu, 
                -{5\over 2}[413]\pi, {5\over 2}[413]\pi$ \\ 
\hline  
\end{tabular}  
\label{table1}  
\end{table*}  
 
The large number of $0^+$ states is difficult to obtain within collective 
models. In Ref. \cite{GCM}, one could obtain at most 5 excited $0^+$ states 
up to 3.2 MeV in calculations with the geometric collective model or  
the interacting boson model (with only $s$- and $d$-boson). 
The reason is that within such collective models, the number of degrees of 
freedom of collective motion is limited. Only if one considered 
the odd-parity bosons in an extended boson space, could the authors in  
Ref. \cite{GCM} get more excited $0^+$ states.   
 
In Table I, we list the 18 calculated $0^+$ states below an excitation 
energy of 3.25 MeV 
(slightly higher than the highest experimental $0^+$ state).  
Their leading  
configurations are also given. Note that each of the configurations 
listed here can not be pure, but is the dominant part in the corresponding 
wave function. One sees that most 2-qp states 
coming from the near-Fermi Nilsson levels have lower energies. 
Due to the varying responses of the residual interactions 
through configuration mixing, one finds that  
some 4-qp states are lower in excitation energy than 2-qp states. 
 
Up to now, we have seen that the PSM calculations,  
which explicitly include two- and four-qp states built 
on the basis but no vibrational degrees of freedom, can reasonably produce  
a sufficient number of $0^+$ states within the right excitation energy range. 
This fact  
suggests that these states are qp states in nature contrary to the work 
of Zamfir, Zhang, and Casten \cite{GCM}. 
However, before we do that, we should study the transition properties 
of these states. 
 
The excited $0^+$ states can decay to the $2^+_g$ state 
in the ground state band 
through E2 transition. One such a transition was measured in Ref.  
\cite{Borner}. Preliminary results of Lesher {\it et. al.} \cite{BE2}  
show 12 such transitions. The PSM  
calculations of the transition from the $i$-th 
$0^+$ state to the $2^+_g$ state,   
B(E2, $0^+_i \rightarrow 2^+_g$), are listed in Table I.  
Comparing the theoretical B(E2) values with the experimental data \cite{BE2}, 
we found that the numbers (in W.u.) listed in Table I are  
one or two orders of 
magnitude too small for many of the transitions.  
 
We note that the B(E2, $0^+_i \rightarrow 2^+_g$) values (a few W.u.) of  
Lesher {\it et. al.} \cite{BE2} are in average of  
two orders of magnitude smaller than the in-band 
B(E2) values of the ground-state band (typically, a few  
hundred W.u.). These small B(E2) values usually indicate  
that the observed $0^+$ states in  
Ref. \cite{Ani02} should carry significant quasiparticle components.   
However, our calculated  
B(E2) results are much smaller than the values of Lesher {\it et. al.},  
which might indicate an insufficient mixing of collectivity to the 
qp states in our model. This tends to suggest that although  
most, if not all,  
of these $0^+$ states have significant qp components,  
they are also mixed with correlations induced by the collective 
motion \cite{QPNM}.   
These correlations can be introduced in the PSM framework  
by inclusion of interactions of higher 
orders of the multipole type \cite{CG01},   
and by addition of the $D$-pair operators to the vacuum state \cite{Heavy},  
which takes both quasiparticle and 
collective degrees of freedom explicitly into account in a shell model 
basis. Generator Coordinate Method, which consists of a construction of
a linear superposition of different product wave functions, can also be
adopted by the PSM. 
 
In summary, the projected shell model was employed to understand the nature 
of excited $0^+$ states in deformed nuclei. The shell model space consists 
of projected 2- and 4-qp states on top of the deformed BCS vacuum state. 
Therefore, the calculation emphasized the quasiparticle character of these 
states, in contrast to previous work \cite{GCM}  
using collective models.   
Our results were compared with the remarkable example of recently observed 
13 excited $0^+$ states in $^{158}$Gd \cite{Exp}.  
After performing exact angular momentum projection and configuration 
mixing calculations 
(by two-body residual interactions)  
for all the possible low-lying qp configurations  
that can give rise to $K^\pi = 0^+$ states,  
we found that the obtained energy levels as well as the number of the states   
can reasonably explain the data.  
Preliminary measurements of B(E2) values suggest that mixing of 
these qp states with the collective, vibrational motion may not be neglected. 
     
Research at University of Notre Dame is supported by the NSF through contract 
No. PHY-0140324.   
                            Research at the University of 
                            Tennessee is supported by the U.~S. Department 
                            of Energy through Contract No.\ 
                            DE--FG05--96ER40983. 
   
\baselineskip = 14pt   
\bibliographystyle{unsrt}   
     
\end{document}